\documentclass[12pt]{article}
\pdfoutput=1

\usepackage{amsmath}
\usepackage{amssymb}
\usepackage{epsfig}
\usepackage{epstopdf}
\usepackage{latexsym}
\usepackage{color}
\numberwithin{equation}{section}
\usepackage[
      colorlinks=false,
      linkcolor=darkblue,  
      urlcolor=blue,    
      filecolor=blue,     
      citecolor=red,
linktocpage=true,
      pdfstartview=FitV,
      bookmarksopen=true    
      ]{hyperref}

\begin{document}

\begin{titlepage}

\centerline
\centerline
\centerline
\bigskip
\centerline{\huge \rm Brane-jet stabilities from Janus}
\bigskip
\centerline{\huge \rm and Sasaki-Einstein}
\bigskip
\bigskip
\bigskip
\bigskip
\bigskip
\bigskip
\bigskip
\bigskip
\centerline{\rm Minwoo Suh}
\bigskip
\centerline{\it Department of Physics, Kyungpook National University, Daegu 41566, Korea}
\bigskip
\centerline{\tt minwoosuh1@gmail.com} 
\bigskip
\bigskip
\bigskip
\bigskip
\bigskip
\bigskip
\bigskip
\bigskip

\begin{abstract}
\noindent We show that there are certain perturbatively stable non-supersymmetric $AdS$ vacua which are also brane-jet stable. Also we extend the analysis of brane-jets to the $AdS$ vacua from curved domain walls like Janus solutions. First, we apply the brane-jet analysis to the non-supersymmetric Janus solutions of type IIB supergravity found by Bak, Gutperle and Hirano. Second, we study the brane-jet of $AdS_4$ vacua from eleven-dimensional supergravity on Sasaki-Einstein manifolds: the supersymmetric and the skew-whiffed Freund-Rubin, the Pope-Warner, and the Englert solutions. Third, we examine the non-supersymmetric $AdS_4$ vacua from $Q^{1,1,1} $ and $M^{1,1,1}$ manifolds discovered by Cassani, Koerber and Varela. It turns out that all the $AdS$ vacua we consider in this work are brane-jet stable. Especially, the Janus, the skew-whipped Freund-Rubin, and the $AdS_4$ vacua from $Q^{1,1,1} $ and $M^{1,1,1}$ are perturbatively stable within known subsectors of truncations and also brane-jet stable.
\end{abstract}

\vskip 3cm

\flushleft {October, 2021}

\end{titlepage}

\tableofcontents

\section{Introduction}

Recently, there has been various developments expanding our understanding of the structure of vacua from string and M-theory. First, conjectures are proposed which restrict the possible landscape of quantum gravity. In particular, a strong version, \cite{Ooguri:2016pdq, Freivogel:2016qwc}, of the weak gravity conjecture, \cite{Arkani-Hamed:2006emk}, implies the non-existence of stable non-supersymmetric $AdS$ vacua in quantum gravity. Second, machine learning techniques enable us to discover a larger landscape of $AdS$ vacua in string and M-theory, \cite{Comsa:2019rcz, Bobev:2019dik, Krishnan:2020sfg, Bobev:2020ttg, Bobev:2020qev}. In gauged supergravity, due to the complexity of scalar potentials, search of critical points was a daunting task. Third, as an application of exceptional field theory, a powerful tool of Kaluza-Klein spectroscopy for mass spectrum was developed in \cite{Malek:2019eaz, Malek:2020mlk, Malek:2020yue, Varela:2020wty, Cesaro:2020soq}. It is useful in checking the perturbative stability of an $AdS$ vacuum by the Breitenlohner-Freedman (BF) bound, \cite{Breitenlohner:1982bm, Breitenlohner:1982jf, Gibbons:1983aq}.

In addition, a new decay channel for $AdS$ vacua, called the brane-jet instability, was proposed in \cite{Bena:2020xxb}. Along the lines of previous studies, \cite{Maldacena:1998uz, Danielsson:2017max, Apruzzi:2019ecr}, it employs a probe brane to test the instability. When the force acting on the probe brane is repulsive, the vacuum is determined to be unstable. In close relation to the developments we mentioned, the brane-jet instabilities of numerous $AdS$ vacua from gauged supergravity in diverse dimensions have been tested, \cite{Bena:2020xxb, Suh:2020rma, Guarino:2020jwv, Apruzzi:2021nle}. In the end, among the non-supersymmetric $AdS$ vacua which have been tested in the literature, only seven $AdS_4$ vacua of massive type IIA supergravity, \cite{Guarino:2015jca, Guarino:2015qaa, Guarino:2015vca}, are proven to be perturbatively stable with a subsector of truncations, \cite{Guarino:2020flh}, and also brane-jet stable, \cite{Guarino:2020jwv}.{\footnote{Recently, in \cite{Bomans:2021ara} so-called $dilaton$ bubble solution was explicitly constructed for the $G_2$-symmetric vacuum which is one of the seven $AdS_4$ vacua of massive type IIA supergravity known to be BF and brane-jet stable.}} 

In this paper, we will show that there are more non-supersymmetric $AdS$ vacua which are both perturbatively stable within known subsectors of truncations and also brane-jet stable: the non-supersymmetric Janus solutions of type IIB supergravity, the skew-whipped Freund-Rubin, and the $AdS_4$ vacua on $Q^{1,1,1} $ and $M^{1,1,1}$ manifolds in eleven-dimensional supergravity. They are possible counter-examples to the swampland conjecture for non-supersymmetric $AdS$ vacua, \cite{Ooguri:2016pdq, Freivogel:2016qwc}.

\subsection{Janus solutions of type IIB supergravity}

So far, the brane-jet instability was only tested for $AdS$ vacua from flat domain walls. We test the brane-jet instability of $AdS$ solutions from curved domain walls, namely the non-supersymmetric Janus solutions of type IIB supergravity, \cite{Schwarz:1983qr, Howe:1983sra}, obtained by Bak, Gutperle and Hirano in \cite{Bak:2003jk}, which is the first Janus solution constructed in the literature. This class of solutions were shown to be stable against the scalar perturbations in \cite{Bak:2003jk} and by the Nester-Witten positive energy theorem in \cite{Freedman:2003ax}.

In order to understand the brane-jet in the curved domain walls, we start by studying a simple example: we put the famous supersymmetric $AdS_5\times{S}^5$ solution in the $AdS_4$-sliced coordinates and compare the brane-jet with the usual $AdS_5\times{S}^5$ solution in the Poincar{\'e} coordinates. In the case of curved domain walls, the worldvolume of probe D3-brane is on $AdS_4$ instead of Mink$_4$ of the usual flat domain walls. As supersymmetric solutions, they are both brane-jet stable, but display different behaviours of brane-jets. Then we study the brane-jet of the non-supersymmetric Janus solutions. The non-supersymmetric Janus solutions turn out to be brane-jet stable.

\subsection{$AdS_4$ vacua from Sasaki-Einstein manifolds}

For a Sasaki-Einstein manifold, $SE_7$, there is the Freund-Rubin solution, \cite{Freund:1980xh}, which is $\mathcal{N}\,=\,2$ supersymmetric $AdS_4\,\times\,SE_7$ solution of eleven-dimensional supergravity, \cite{Cremmer:1978km}. There are also non-supersymmetric solutions: the skew-whiffed Freund-Rubin, the Pope-Warner, and the Englert solutions.

The skew-whiffed Freund-Rubin solutions, \cite{Freund:1980xh}, are obtained by flipping the orientation of $SE_7$ of supersymmetric Freund-Rubin solutions. This solution breaks all the supersymmetry. However, if $SE_7$ is $S^7$, it preserves the full supersymmetry. As this solution inherits its mass spectrum from the superymmetric one, it is automatically perturbatively stable by the Breitenlohner-Freedman bound, \cite{Duff:1984sv}.

The Pope-Warner solutions, \cite{Pope:1984bd, Pope:1984jj}, break all the supersymmetry. However, they were shown to be stable within the massive truncations of \cite{Gauntlett:2009zw, Gauntlett:2009bh, Cassani:2012pj}. On the other hand, when $SE_7$ is $S^7$, unstable modes were found in the $SU(4)^-$-invariant sector of gauged $\mathcal{N}\,=\,8$ supergravity, \cite{deWit:1982bul}, which is the consistent truncation on the solutions, \cite{Bobev:2010ib}. They were also proved to be unstable within the massive truncation on tri-Sasakian manifolds, \cite{Cassani:2011fu}. The final perturbative instability of these solutions on arbitrary Sasaki-Einstein manifolds was proved in \cite{Pilch:2013gda}.

The Englert solutions were first found on $S^7$ in \cite{Englert:1982vs}, and were generalized to $SE_7$ in \cite{Awada:1982pk}. The Englert solutions break all the supersymmetry, \cite{Englert:1983qe, Page:1984fu}. The perturbatively unstable modes of the solutions were found in \cite{Page:1984fu} and also within the massive truncations of \cite{Gauntlett:2009zw, Gauntlett:2009bh}. This solution corresponds to the $SO(7)^-$ fixed point of gauged $\mathcal{N}\,=\,8$ supergravity and the instability of the fixed point was shown in \cite{deWit:1983gs, Biran:1984jr}.

To recapitulate, among the non-supersymmetric $AdS_4$ solutions from Sasaki-Einstein manifolds, the skew-whiffed Freund-Rubin solutions are known to be perturbatively stable and the Pope-Warner and the Englert solutions are proved to be unstable. For a more detailed account on the stability analysis of the solutions, see the introduction of \cite{Pilch:2013gda} and section 3 of \cite{Gauntlett:2009bh}.

Additionally, there are non-supersymmetric $AdS_4$ solutions from eleven-dimensional supergravity on Sasaki-Einstein manifolds of $Q^{1,1,1} $ and $M^{1,1,1}$ found by Cassani, Koerber and Varela in \cite{Cassani:2012pj}. These solutions were proven to be perturbatively stable within the truncation performed there.

So far we have reviewed a number of $AdS_4\,\times\,SE_7$ solutions in eleven-dimensional supergravity. In this paper, we will examine the brane-jet instabilities of these solutions. It turns out that all solutions we consider are brane-jet stable. Thus the skew-whiffed Freund-Rubin solutions and the solutions on $Q^{1,1,1} $ and $M^{1,1,1}$ are both perturbatively and brane-jet stable. Other $AdS_4$ vacua are already known to be perturbatively unstable.

\bigskip

In section 2, we study the brane-jet of the non-supersymmetric Janus solutions. In section 3, we consider the supersymmetric Freund-Rubin, the skew-whiffed Freund-Rubin, the Pope-Warner, and the Englert solutions and calculate the M2-brane probe potentials to examine the brane-jet instability. In section 4, we consider the solutions from $Q^{1,1,1} $ and $M^{1,1,1}$ manifolds and calculate the M2-brane probe potentials to examine the brane-jet instability. We conclude in section 5. In an appendix, we present the normalization of the supersymmetric $AdS_5\times{S}^5$ solution of type IIB supergravity.

\section{Non-supersymmetric Janus solutions}

\subsection{The solutions}

We review the non-supersymmetric Janus solutions of type IIB supergravity in \cite{Bak:2003jk}. The solutions in the Einstein frame are given by the metric, the dilaton field, and the five-form flux, respectively,
\begin{align} \label{janusjanus}
ds^2\,=&\,f(r)\left(ds_{AdS_4}^2+dr^2\right)+ds_{S^5}^2\,, \notag \\
\phi\,=&\,\phi(r)\,, \notag \\
F_{(5)}\,=&\,4f(r)^{5/2}dr\wedge\,\text{vol}_{AdS_4}+4\text{vol}_{S^5}\,,
\end{align}
where the range of $r$ is $r\in[-\frac{\pi}{2},\frac{\pi}{2}]$.{\footnote{We corrected the factors in the five-form flux to be 4 instead of 2 in \cite{Bak:2003jk}. See appendix A for the normalization of the $AdS_5\times{S}^5$ solutions.}}

The Einstein equations give
\begin{align}
2f'f'-2ff''\,=&\,-4f^3+\frac{1}{2}f^2\phi'\phi'\,, \notag \\
12f^2+f'f'+2ff''\,=&\,16f^3\,.
\end{align}
The equation of motion for the dilaton field reduces to
\begin{equation}
\phi'(r)\,=\,\frac{c_0}{f(r)^{3/2}}\,,
\end{equation}
where $c_0$ is a constant.  We obtain numerical solutions and present them in Figure 1.
\begin{figure}[t]
\begin{center}
\includegraphics[width=3.0in]{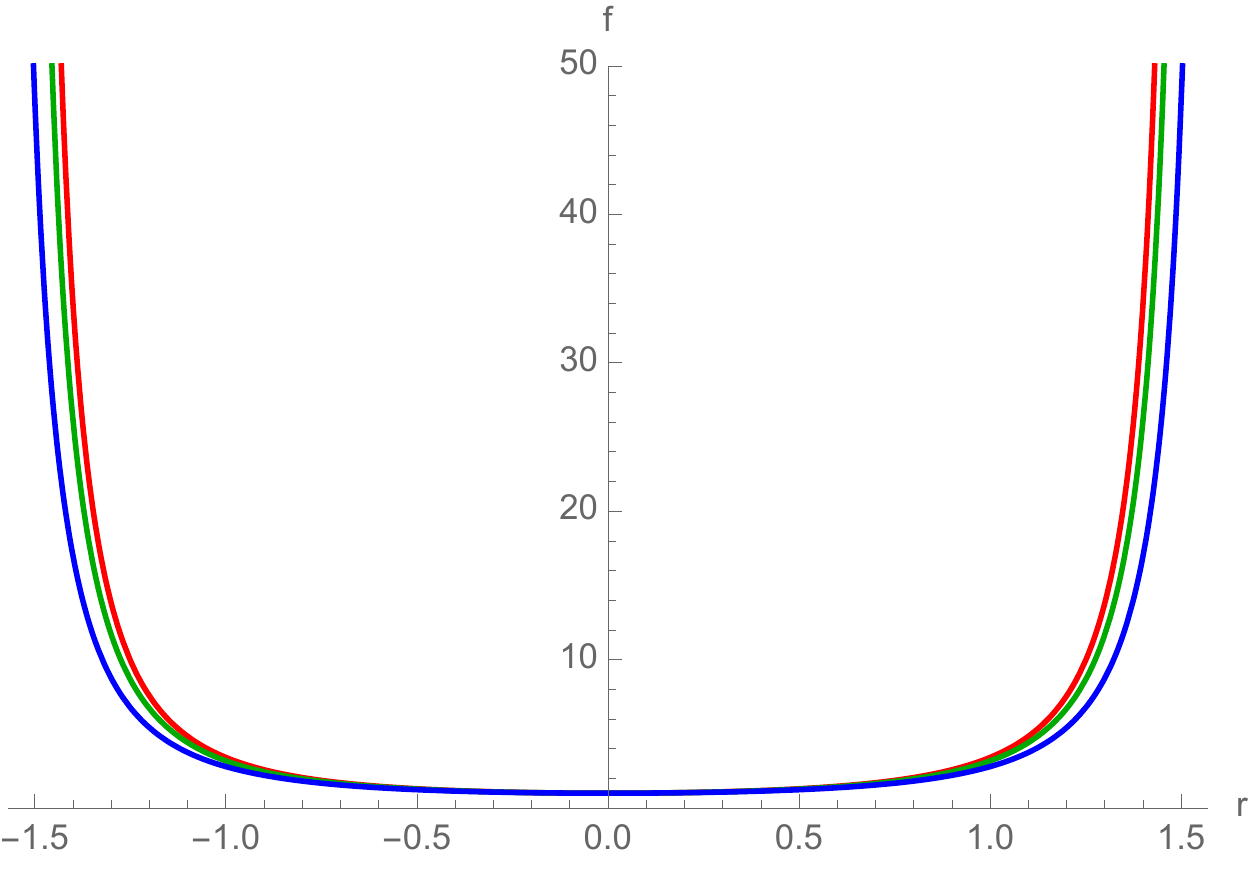} \qquad \includegraphics[width=3.0in]{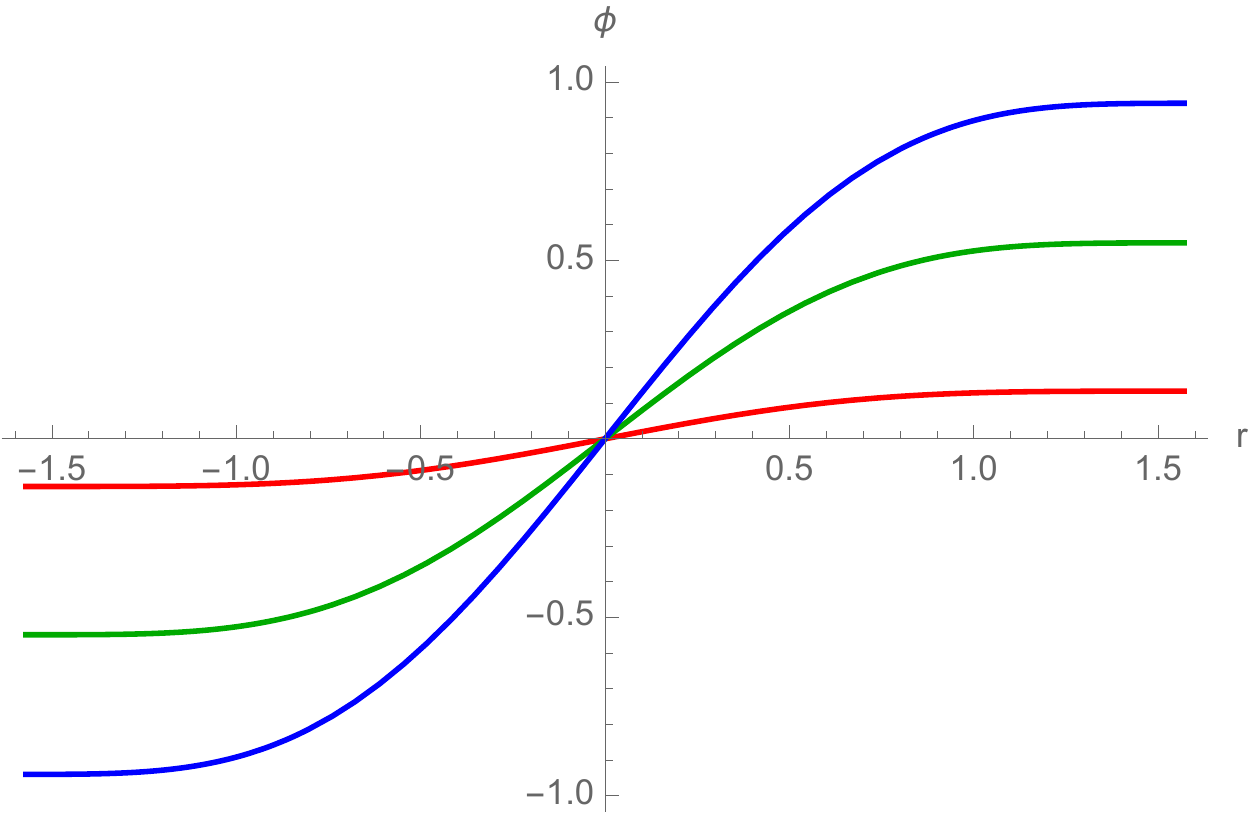}
\caption{{\it Representative solutions for $f(r)$ and $\phi(r)$ with $c_0\,=\,0.2,\,0.8,\,1.3$ in red, green, and blue, respectively.}}
\label{1}
\end{center}
\end{figure}

When we have $c_0=0$ and
\begin{equation} \label{tosj}
f(r)\,=\,\frac{1}{\cos^2r}\,.
\end{equation}
the solutions reduces to the supersymmetric $AdS_5\times{S}^5$ solution in the $AdS_4$-sliced coordinates.

\subsection{D3-brane probe}

\subsubsection{The $AdS_5\times{S}^5$ solution}

We consider the supersymmetric $AdS_5\times{S}^5$ solution of type IIB supergravity. The metric and the self-dual five-form flux are given, respectively, by
\begin{align}
ds^2\,=&\,ds_{AdS_5}^2+ds_{S^5}^2\,, \notag \\
F_{(5)}\,=&\,4\text{vol}_{AdS_5}+4\text{vol}_{S^5}\,.
\end{align}
We take the Poincar{\'e} coordinates for the $AdS_5$ and the corresponding volume form for the flux,
\begin{align}
ds^2\,=&\,\frac{-dt^2+dx^2+dy^2+dz^2+dr^2}{r^2}+ds^2_{S^5}\,, \notag \\
F_{(5)}\,=&\,\frac{4}{r^5}dt\wedge{d}x\wedge{d}y\wedge{d}z\wedge{d}r+4\text{vol}_{S^5}\,.
\end{align}
Then the four-form potential is given by
\begin{equation}
C_{(4)}\,=\,-\frac{1}{r^4}dt\wedge{d}x\wedge{d}y\wedge{d}z+\ldots\,.
\end{equation}

We partition the spacetime coordinates,
\begin{equation}
x^a\,=\,\{t,x,y,z\}\,, \qquad y^m\,=\,\{r,\ldots\}\,,
\end{equation}
and choose the static gauge,
\begin{equation}
x^a\,=\,\xi^a\,, \qquad y^m\,=\,y^m(t)\,,
\end{equation}
where $\xi^a$ are the worldvolume coordinates. The pull-back of the metric is
\begin{equation}
\tilde{G}_{ab}\,=\,G_{\mu\nu}\frac{\partial{x}^\mu}{\partial\xi^a}\frac{\partial{x}^\nu}{\partial\xi^b}\,.
\end{equation}

Now we study the worldvolume action of the D3-brane which is given by a sum of DBI and WZ terms. If the probe brane moves slowly, the worldvolume action in the Einstein frame is, $e.g.$, \cite{Clark:2004sb},
\begin{align}
S\,=&\,-\int{d}^4\xi\sqrt{-\text{det}(\tilde{G})}-\int\tilde{C}_{(4)} \notag \\
=&\,-\int{d}^4\xi\left(\frac{1}{r^4}-\frac{1}{2r^2}G_{mn}\dot{y}^m\dot{y}^n\right)-\int\left(-\frac{1}{r^4}\right)dt\wedge\,dx\wedge\,dy\wedge\,dz\,,
\end{align}
where $\tilde{C}_{(4)}$ is the pull-back of the four-form potential. Then the worldvolume action reduces to
\begin{equation}
S\,=\,\int{d}^4\xi\left(K-V\right)\,,
\end{equation}
where the kinetic and the potential terms are
\begin{align}
K\,=&\,\frac{1}{2r^2}G_{mn}\dot{y}^m\dot{y}^n+\ldots\,, \notag \\
V\,=&\,\frac{1}{r^4}-\frac{1}{r^4}\,=\,0\,.
\end{align}
For the $AdS_5\times{S}^5$ solution, potential for the probe D3-brane vanishes identically. It is non-negative and, thus, brane-jet stable as we expect for supersymmetric solutions.

\subsubsection{The $AdS_5\times{S}^5$ solution in the $AdS_4$-sliced coordinates}

We consider the $AdS_5\times{S}^5$ solution in the $AdS_4$-sliced coordinates, \eqref{tosj}, \cite{Bak:2003jk}. The metric and the self-dual five-form flux are given, respectively, by
\begin{align}
ds^2\,=&\,f(r)\left(ds_{AdS_4}^2+dr^2\right)+ds_{S^5}^2\,, \notag \\
F_{(5)}\,=&\,4f(r)^{5/2}dr\wedge\,\text{vol}_{AdS_4}+4\text{vol}_{S^5}\,,
\end{align}
where the function, $f(r)$, is given by
\begin{equation}
f(r)\,=\,\frac{1}{\cos^2r}\,.
\end{equation}
We take the Poincar{\'e} coordinates for the $AdS_4$,
\begin{align}
ds^2_{AdS_4}\,=&\,\frac{-dt^2+dx^2+dy^2+dz^2}{z^2}\,.
\end{align}
The four-form potential is given by
\begin{equation}
C_{(4)}\,=\,\left(4\int{f}(r)^{5/2}dr\right)\text{vol}_{AdS_4}+\ldots\,.
\end{equation}

We partition the spacetime coordinates,
\begin{equation}
x^a\,=\,\{t,x,y,z\}\,, \qquad y^m\,=\,\{r,\ldots\}\,,
\end{equation}
and choose the static gauge,
\begin{equation}
x^a\,=\,\xi^a\,, \qquad y^m\,=\,y^m(t)\,,
\end{equation}
where $\xi^a$ are the worldvolume coordinates. The pull-back of the metric is
\begin{equation}
\tilde{G}_{ab}\,=\,G_{\mu\nu}\frac{\partial{x}^\mu}{\partial\xi^a}\frac{\partial{x}^\nu}{\partial\xi^b}\,.
\end{equation}

\begin{figure}[t]
\begin{center}
\includegraphics[width=3in]{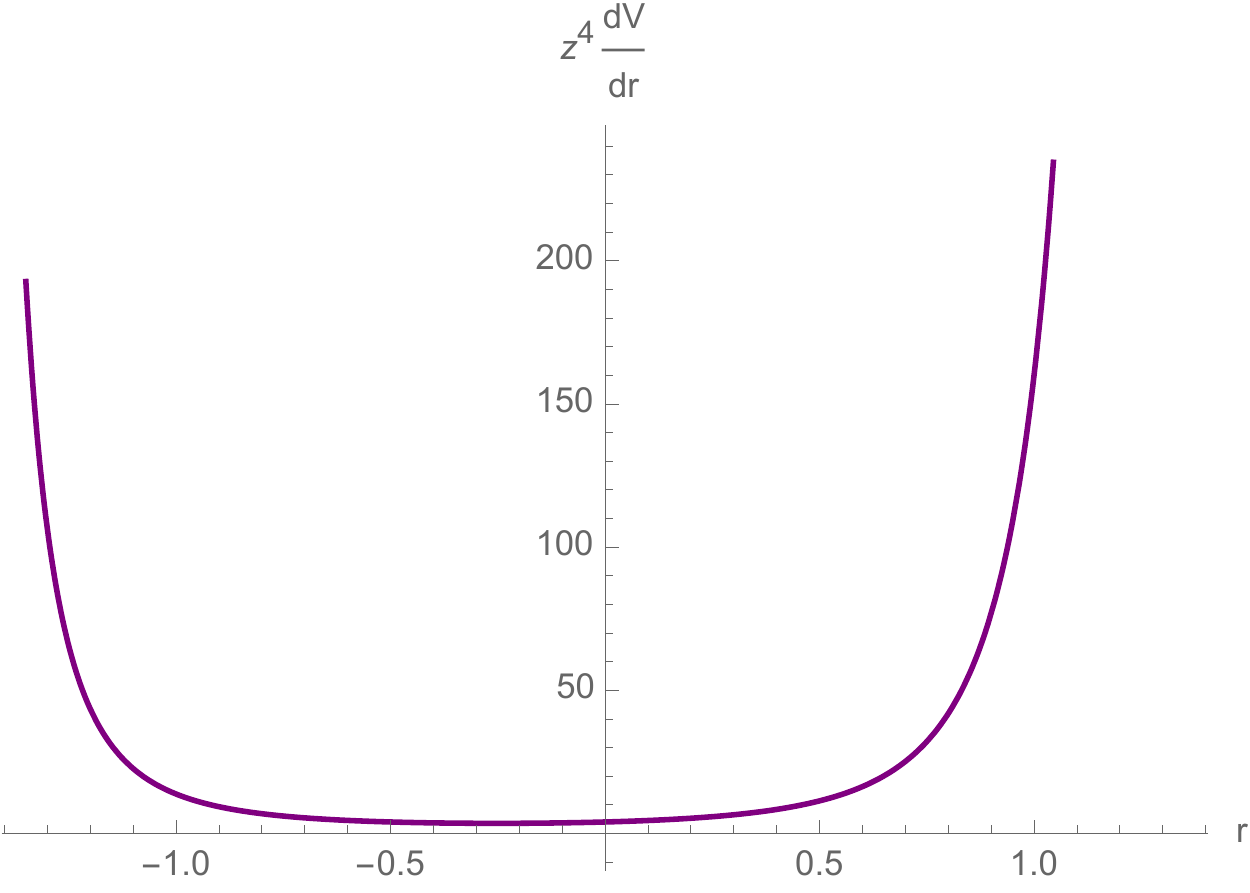}
\caption{{\it The force acting on probe D3-brane for the $AdS_5\times{S}^5$ solution in the $AdS_4$-sliced coordinates.}}
\label{1}
\end{center}
\end{figure}

Now we study the worldvolume action of the D3-brane which is given by a sum of DBI and WZ terms. If the probe brane moves slowly, the worldvolume action in the Einstein frame is, $e.g.$, \cite{Clark:2004sb},
\begin{align}
S\,=&\,-\int{d}^4\xi\sqrt{-\text{det}(\tilde{G})}-\int\tilde{C}_{(4)} \notag \\
=&\,-\int{d}^4\xi\frac{1}{z^4}\left(f^2-\frac{f}{2}G_{mn}\dot{y}^m\dot{y}^n\right)-\int\frac{1}{z^4}\left(4\int{f}^{5/2}dr\right)dt\wedge\,dx\wedge\,dy\wedge\,dz\,,
\end{align}
where $\tilde{C}_{(4)}$ is the pull-back of the four-form potential. Then the worldvolume action reduces to
\begin{equation}
S\,=\,\int{d}^4\xi\left(K-V\right)\,,
\end{equation}
where the kinetic and the potential terms are
\begin{align}
K\,=&\,\frac{1}{z^4}\frac{f}{2}G_{mn}\dot{y}^m\dot{y}^n+\ldots\,, \notag \\
V\,=&\,\frac{1}{z^4}\left(f^2+4\int{f}^{5/2}dr\right)\,.
\end{align}
We numerically compute the force acting on the probe D3-brane, $z^4\frac{dV}{dr}$, and present it in Figure 2. It is positive and the force is attractive. Thus it is brane-jet stable as we expect for supersymmetric solutions. Furthermore, as the force reaches the minimum value of 3.52528 around $r=-0.252$ and never vanishes, there is no equilibrium point.

\subsubsection{The non-supersymmetric Janus solutions}

We consider the non-supersymmetric Janus solutions presented in \eqref{janusjanus}. We partition the spacetime coordinates,
\begin{equation}
x^a\,=\,\{t,x,y,z\}\,, \qquad y^m\,=\,\{r,\ldots\}\,,
\end{equation}
and choose the static gauge,
\begin{equation}
x^a\,=\,\xi^a\,, \qquad y^m\,=\,y^m(t)\,,
\end{equation}
where $\xi^a$ are the worldvolume coordinates. The pull-back of the metric is
\begin{equation}
\tilde{G}_{ab}\,=\,G_{\mu\nu}\frac{\partial{x}^\mu}{\partial\xi^a}\frac{\partial{x}^\nu}{\partial\xi^b}\,.
\end{equation}

\begin{figure}[t]
\begin{center}
\includegraphics[width=3in]{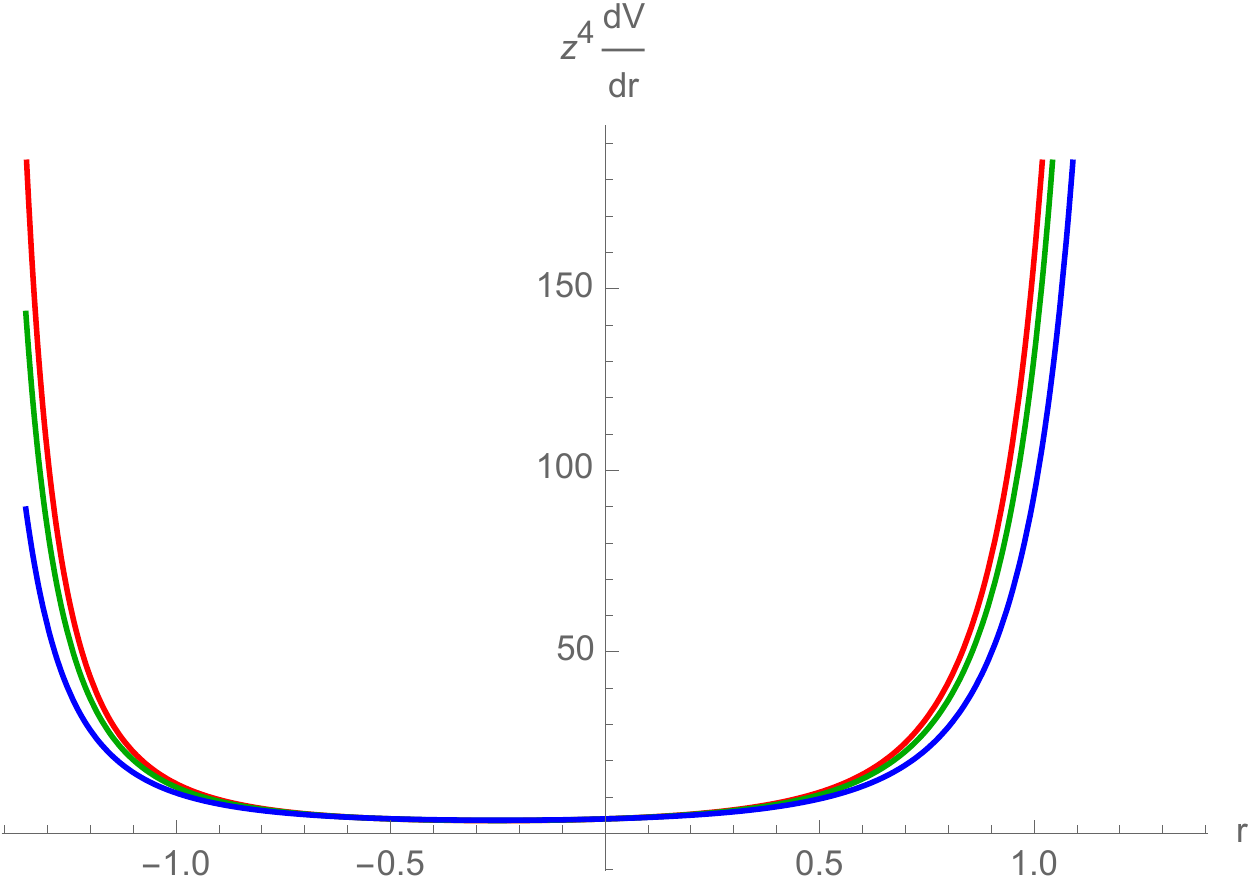}
\caption{{\it The force acting on probe D3-brane for the non-supersymmetric Janus solutions with $c_0\,=\,0.2,\,0.8,\,1.3$ in red, green, and blue, respectively.}}
\label{1}
\end{center}
\end{figure}

Now we study the worldvolume action of the D3-branes which is given by a sum of DBI and WZ terms. If the probe branes move slowly, the worldvolume action in the Einstein frame is, $e.g.$, \cite{Clark:2004sb},
\begin{align}
S\,=&\,-\int{d}^4\xi\,\sqrt{-\text{det}(\tilde{G})}-\int\tilde{C}_{(4)} \notag \\
=&\,-\int{d}^4\xi\frac{1}{z^4}\left(f^2-\frac{f}{2}G_{mn}\dot{y}^m\dot{y}^n\right)-\int\frac{1}{z^4}\left(4\int{f}^{5/2}dr\right)dt\wedge\,dx\wedge\,dy\wedge\,dz\,,
\end{align}
where $\tilde{C}_{(4)}$ is the pull-back of the four-form potential. Then the worldvolume action reduces to
\begin{equation}
S\,=\,\int{d}^4\xi\left(K-V\right)\,,
\end{equation}
where the kinetic and the potential terms are
\begin{align}
K\,=&\,\frac{1}{z^4}\frac{f}{2}G_{mn}\dot{y}^m\dot{y}^n+\ldots\,, \notag \\
V\,=&\,\frac{1}{z^4}\left(f^2+4\int{f}^{5/2}dr\right)\,.
\end{align}
We numerically compute the force acting on the probe D3-brane, $z^4\frac{dV}{dr}$, and present it in Figure 3. It is positive and the force is attractive. Thus we conclude that the non-supersymmetric Janus solutions are brane-jet stable. Furthermore, as the force reaches the minimum value of 3.52786 around $r=-0.253$ and never vanishes, there is no equilibrium point.

\section{$AdS_4$ vacua from arbitrary Sasaki-Einstein manifolds}

\subsection{Freund-Rubin, skew-whiffed, Pope-Warner, and Englert solutions}

We consider $AdS_4$ solutions of eleven-dimensional supergravity on arbitrary seven-dimensional Sasaki-Einstein manifolds. In particular, we review the supersymmetric Freund-Rubin, the skew-whiffed Freund-Rubin, the Pope-Warner, and the Englert solutions. To present the solutions in a uniform manner, we employ the ansatz used for the consistent truncation of eleven-dimensional supergravity on arbitrary seven-dimensional Sasaki-Einstein manifolds, \cite{Gauntlett:2009zw, Gauntlett:2009bh}.

Locally Sasaki-Einstein manifold is a fibration over a K\"ahler-Einstein manifold,
\begin{equation}
ds^2_{SE_7}\,=\,ds^2_{KE_6}+\eta\otimes\eta\,,
\end{equation}
where $\eta$ is the one-form dual to the Reeb Killing vector from $d\eta\,=\,2J$ and $J$ is the K\"ahler form of $KE_6$. The (3,0)-form on $KE_6$ is denoted by $\omega$ and satisfies $d\Omega\,=\,4i\eta\wedge\Omega$. Then the volume form is $vol_{SE_7}\,=\,\eta\wedge{J}^3/3!\,=\,(i/8)\eta\wedge\Omega\wedge\Omega^*$.

The metric employed for the consistent truncation on general Sasaki-Einstein manifolds is given by, \cite{Gauntlett:2009bh},
\begin{equation}
\frac{1}{(2L)^2}ds^2\,=\,e^{-6U-V}ds_4^2+e^{2U}ds^2_{KE_6}+e^{2V}\left(\eta+A_1\right)\otimes\left(\eta+A_1\right)\,,
\end{equation}
and the four-form flux is
\begin{align}
\frac{1}{(2L)^3}G_4\,=\,6&e^{-18U-3V}\left(\epsilon+h^2+|\chi|^2\right)vol_4+H_3\wedge\left(\eta+A_1\right)+H_2\wedge{J} \notag \\
&+dh\wedge{J}\wedge\left(\eta+A_1\right)+2hJ\wedge{J} \notag \\
&+\sqrt{3}\left[\chi\left(\eta+A_1\right)\wedge\Omega-\frac{i}{4}D\chi\wedge\Omega+c.c.\right]\,,
\end{align}
where $\epsilon\,=\,\pm{1}$, $D\chi\,=\,d\chi-4iA_1\chi$ and $L$ is an overall scale parameter. $U$, $V$, $h$ are real scalar fields and $\chi$ is a complex scalar field in four dimensions. In four dimensions there are also one- and two-form fields, $A_1$, $B_1$ and $B_2$, with field strengthes,
\begin{align}
F_2\,=&\,dA_1\,, \notag \\
H_3\,=&\,dB_2\,, \notag \\
H_2\,=&\,dB_1+2B_2+hF_2\,.
\end{align}

There are previously known $AdS_4\,\times\,SE_7$ solutions.{\footnote{In the consistent truncation to four-dimensional gauged supergravity, \cite{Gauntlett:2009zw, Gauntlett:2009bh}, the solutions we consider are fixed points of the scalar potential,
\begin{equation}
\mathcal{P}\,=\,48e^{-8U-V}-6e^{-10U+V}-24h^2e^{-14U-V}-18\left(\epsilon+h^2+|\chi|^2\right)^2e^{-18U-3V}-24e^{-12U-3V}|\chi|^2\,.
\end{equation}}} The supersymmetric Freund-Rubin solution, \cite{Freund:1980xh}, is
\begin{equation} \label{frs}
\epsilon\,=\,+1\,, \qquad U\,=\,0\,, \qquad V\,=\,0\,, \qquad \chi\,=\,0\,, \qquad h\,=\,0\,, \qquad R^2_{AdS_4}\,=\,\frac{1}{4}\,,
\end{equation}
and is explicitly given by
\begin{align}
\frac{1}{(2L)^2}ds^2\,&=\,\frac{1}{4}ds^2_{AdS_4}+ds^2_{SE_7}\,, \notag \\
\frac{1}{(2L)^3}G_4\,&=\,\epsilon\frac{3}{8}vol_{AdS_4}\,.
\end{align}
Flipping the sign of the four-form flux by choosing $\epsilon\,=\,-1$, we obtain the skew-whiffed Freund-Rubin solution which breaks all the supersymmetry.

The Pope-Warner solution, \cite{Pope:1984bd, Pope:1984jj}, is
\begin{equation} \label{pws}
\epsilon\,=\,-1\,, \qquad e^U\,=\,2^{-1/6}\,, \qquad e^V\,=\,2^{1/3}\,, \qquad \chi^2\,=\,2/3\,, \qquad h\,=\,0\,, \qquad R^2_{AdS_4}\,=\,\frac{3}{16}\,,
\end{equation}
and is explicitly given by
\begin{align}
\frac{1}{(2L)^2}ds^2\,&=\,2^{2/3}\left[\frac{3}{16}ds^2_{AdS_4}+\frac{1}{2}ds^2_{KE_6}+\eta\otimes\eta\right]\,, \notag \\
\frac{1}{(2L)^3}G_4\,&=\,2\left[-\frac{9}{64}vol_{AdS_4}+\frac{1}{\sqrt{2}}\left(\eta\wedge\Omega+c.c.\right)\right]\,.
\end{align}
This solution breaks all the supersymmetry.

The Englert solution, \cite{Englert:1982vs}, is
\begin{equation} \label{es}
\epsilon\,=\,-1\,, \,\,\,\, e^U\,=\,(4/5)^{1/6}\,, \,\,\,\, e^V\,=\,(4/5)^{1/6}\,, \,\,\,\, \chi^2\,=\,4/15\,, \,\,\,\, h^2\,=\,1/5\,, \,\,\, R^2_{AdS_4}\,=\,\frac{12}{25\sqrt{5}}\,,
\end{equation}
and is explicitly given by
\begin{align}
\frac{1}{(2L)^2}ds^2\,&=\,\left(\frac{4}{5}\right)^{1/3}\left[\frac{3}{10}ds^2_{AdS_4}+ds^2_{KE_6}+\eta\otimes\eta\right]\,, \notag \\
\frac{1}{(2L)^3}G_4\,&=\,\left(\frac{4}{5}\right)^{1/2}\left[-\frac{9}{25}vol_{AdS_4}+J\wedge{J}+\left(\eta\wedge\Omega+c.c.\right)\right]\,.
\end{align}
This solution also breaks all the supersymmetry.

\subsection{M2-brane probe}

At the $AdS_4$ fixed points, we have
\begin{equation}
ds^2_4\,=\,e^{2A}\left(-dx_0^2+dx_1^2+dx_2^2\right)+dr^2\,,
\end{equation}
where
\begin{equation}
A\,=\,\frac{r}{l}\,,
\end{equation}
and $l$ is the radius of $AdS_4$. 
We obtain that the three-form potential is
\begin{equation}
\frac{1}{(2L)^3}A_3\,=\,\frac{l}{3}e^{3A}6e^{-18U-3V}\left(\epsilon+h^2+|\chi|^2\right)dx_0\wedge{d}x_1\wedge{d}x_2+\ldots\,,
\end{equation}

We partition the spacetime coordinates,
\begin{equation}
x^a\,=\,\{x_0,x_1,x_2\}\,, \qquad y^m\,=\,\{r,\ldots\}\,,
\end{equation}
and choose the static gauge,
\begin{equation}
x_0\,=\,t\,=\,\xi^0\,,\qquad x^a\,=\,\xi^a\,, \qquad y^m\,=\,y^m(t)\,,
\end{equation}
where $\xi^a$ are the worldvolume coordinates. The pull-back of the metric is
\begin{equation}
\tilde{G}_{ab}\,=\,G_{\mu\nu}\frac{\partial{x}^\mu}{\partial\xi^a}\frac{\partial{x}^\nu}{\partial\xi^b}\,.
\end{equation}

Now we study the worldvolume action of the M2-branes and anti-M2-branes which is given by a sum of DBI and WZ terms. If the probe branes move slowly, the worldvolume action is
\begin{align}
S\,=&\,-\int{d}^3\xi\sqrt{-\text{det}(\tilde{G})}\pm\int\tilde{A}_3 \notag \\
=&\,-\left(2L\right)^3\int{d}^3\xi\left(e^{-9U-3V/2+3A}-\frac{1}{2}e^{-3U-V/2+A}G_{mn}\dot{y}^m\dot{y}^n+\ldots\right) \notag \\
&\,\pm\left(2L\right)^3\int\frac{l}{3}{e}^{3A}6e^{-18U-3V}\left(\epsilon+h^2+|\chi|^2\right)\,{d}x_0\wedge{d}x_1\wedge{d}x_2\,,
\end{align}
where $\tilde{A}_3$ is the pull-back of the three-form potential. Among the $\pm$ signs, the plus and minus signs correspond to the M2-brane and anti-M2-brane probes, respectively. Then the worldvolume action reduces to
\begin{equation}
S\,=\,\left(2L\right)^3\int{d}^3\eta\left(K-\mathcal{V}^\pm\right)\,,
\end{equation}
where the kinetic and the potential terms are
\begin{align}
K\,=&\,\frac{1}{2}e^{-3U-V/2+A}G_{mn}\dot{y}^m\dot{y}^n+\ldots\,, \notag \\
\mathcal{V}^\pm\,=&\,e^{3A}\left(e^{-9U-3V/2}\pm\frac{l}{3}6e^{-18U-3V}\left(\epsilon+h^2+|\chi|^2\right)\right)\,.
\end{align}
For the $AdS_4$ solutions of the supersymmetric Freund-Rubin, the skew-whiffed Freund Rubin, the Pope-Warner, and the Englert solutions in \eqref{frs}, \eqref{pws}, and \eqref{es}, respectively, we obtain
\begin{align} \label{321}
e^{-3A}\mathcal{V}^+|_{\text{SUSY}}\,=&\,0\,, \notag \\
e^{-3A}\mathcal{V}^-|_{\text{skew-whiffed}}\,=&\,0\,, \notag \\
e^{-3A}\mathcal{V}^+|_{\text{Pope-Warner}}\,=&\,2+\frac{2}{\sqrt{3}}\,, \notag \\
e^{-3A}\mathcal{V}^+|_{\text{Englert}}\,=&\,\frac{5^{5/4}}{4\sqrt{3}}+\frac{5^{7/4}}{8\sqrt{2}}\,.
\end{align}
Like the supersymmetric solution, the skew-whiffed solution also has a vanishing force from the anti-M2-brane probe.{\footnote{We would like to thank an anonymous referee who suggested the anti-M2-brane probe.}} From the M2-brane probe, we find $e^{-3A}\mathcal{V}^+|_{\text{skew-whiffed}}\,=\,2$. All M2-brane probe potentials obtained here for non-supersymmetric solutions are non-negative. Thus the force acting on the probe M2-brane, $d\mathcal{V}/dr$, is positive and attractive. We conclude that all solutions we consider are brane-jet stable. However, the Pope-Warner and the Englert solutions are known to be BF unstable. On the other hand, the skew-whiffed solutions are BF and also brane-jet stable.

\section{$AdS_4$ vacua from particular Sasaki-Einstein manifolds}

\subsection{Non-supersymmetric $AdS_4$ solutions on $Q^{1,1,1}$ and $M^{1,1,1}$}

In this section, we consider the non-supersymmetric $AdS_4$ solutions found from the consistent truncation of eleven-dimensional supergravity on seven-dimensional homogeneous Sasaki-Einstein manifolds, specifically on $Q^{1,1,1}$ and $M^{1,1,1}$, \cite{Cassani:2012pj}.

We review gauged $\mathcal{N}\,=\,2$ supergravity in four dimensions from the consistent truncation of eleven-dimensional supergravity on $Q^{1,1,1}$ manifolds, \cite{Cassani:2012pj}. The truncation on $M^{1,1,1}$ manifolds are obtained from the truncation on $Q^{1,1,1}$ by identifying $t^3\,=\,t^1$. The field content consists of 1 gravity multiplet, $\{g_{\mu\nu}, A^0_\mu\}$, 3 vector multiplets, $\{A^i_\mu, t^i\}$, and 1 hypermultiplet, $\{\phi, a, \xi^0, \tilde{\xi}_0\}$, where $i\,=\,1,2,3$. There are 4 real scalar fields, $\phi, a, \xi^0, \tilde{\xi}_0$, where $\phi$ and $a$ are dilaton and axion fields in four dimensions. There are 3 complex scalar fields, $t^i$, for which we also employ the parametrizations,
\begin{equation}
t^i\,=\,b^i+iv^i\,,
\end{equation}
and
\begin{equation}
v^i\,=\,e^{2u_i}\,.
\end{equation}
The scalar fields from the vector multiplets and the hypermultiplet parametrize the coset manifolds,
\begin{equation}
\mathcal{M}_v\times\mathcal{M}_h\,=\,\left(\frac{SU(1,1)}{U(1)}\right)^3\,\times\,\frac{SU(2,1)}{S(U(2)\,\times\,U(1))}\,,
\end{equation}
which is a product of special K\"ahler and quaternionic manifolds, respectively. The metric of the special K\"ahler and quaternionic manifolds are, respectively, given by
\begin{equation}
ds^2\,=\,\sum_{i=1}^3\left[\left(du_i\right)^2+\frac{1}{4}e^{-2u_i}\left(db^i\right)^2\right]\,,
\end{equation}
and
\begin{equation}
h_{uv}dq^udq^v\,=\,\left(d\phi\right)^2+\frac{1}{4}e^{4\phi}\left[da+\frac{1}{2}\left(\xi^0d\tilde{\xi}_0-\tilde{\xi}_0d\xi^0\right)\right]^2+\frac{1}{4}e^{2\phi}\left(d\xi^0\right)^2+\frac{1}{4}e^{2\phi}(d\tilde{\xi}_0)^2\,.
\end{equation}
The scalar potential is given by
\begin{align}
\mathcal{P}\,=\,&-8e^{2\phi}\left(e^{-2u_1}+e^{-2u_2}+e^{-2u_3}\right)+e^{4\phi}\left(e^{-2u_1+2u_2+2u_3}+e^{2u_1-2u_2+2u_3}+e^{2u_1+2u_2-2u_3}\right) \notag \\
&+e^{4\phi-2u_1-2u_2-2u_3}\left[e^{4u_1}\left(b^2+b^3\right)^2+e^{4u_2}\left(b^1+b^3\right)^2+e^{4u_3}\left(b^1+b^2\right)^2\right] \notag \\
&+\frac{1}{4}e^{4\phi-2u_1-2u_2-2u_3}\left[e_0+2b^1b^2+2b^1b^3+2b^2b^3+2\left(\xi^0\right)^2+2(\tilde{\xi}_0)^2\right]^2 \notag \\
&+4e^{4\phi-2u_1-2u_2-2u_3}\left(\left(\xi^0\right)^2+(\tilde{\xi}_0)^2\right)\,.
\end{align}
There is a supersymmetric fixed point of the scalar potential which corresponds to the $AdS_4\,\times\,Q^{1,1,1}$ solution of eleven-dimensional supergravity, $e.g.$, in (3.16) of \cite{Halmagyi:2013sla},
\begin{align} \label{susyfp}
&v^i\,=\,\sqrt{\frac{e_0}{6}}\,, \qquad e^{-2\phi}\,=\,\frac{e_0}{6}\,,\qquad R_{AdS_4}\,=\,\frac{1}{2}\left(\frac{e_0}{6}\right)^{3/4}\,, \notag \\
&e^{3V}\,=\,\sqrt{\frac{e_0}{6}}\,, \qquad b^i\,=\,0\,, \qquad \xi^0\,=\,\tilde{\xi}_0\,=\,0\,.
\end{align}
There is also a non-supersymmetric fixed point first found in (5.10) of \cite{Cassani:2012pj},
\begin{align} \label{nonsusyfp}
&e^{2U_1}\,=\,\left(\frac{9}{5}\right)^{1/3}a\,, \qquad e^{2U_2}\,=\,\left(\frac{9}{5}\right)^{1/3}a\,, \qquad  e^{2U_3}\,=\,\left(\frac{9}{5}\right)^{-2/3}a\,, \qquad e^{2V}\,=\,\frac{2}{7}15^{2/3}a\,, \notag \\
&b^1\,=\,b^2\,=\,b^3\,=\,0\,, \qquad \left(\xi^0\right)^2+(\tilde{\xi}_0)^2\,=\,\frac{172}{49}a^3\,, \qquad e_0\,=\,-\frac{540}{49}a^3\,, 
\end{align}
where $a>0$ is a free parameter. This point is BF-stable within the truncation.

The supersymmetric and non-supersymmetric fixed points of $AdS_4\,\times\,M^{1,1,1}$ are obtained from \eqref{susyfp} and \eqref{nonsusyfp} as particualr cases of $t^3\,=\,t^1$ and $U_2\,=\,U_1$. 

Now we present the uplift formula to eleven-dimensional supergravity. The metric is given by
\begin{equation}
ds^2\,=\,e^{2V}\mathcal{K}^{-1}ds_4^2+e^{-V}ds^2(B_6)+e^{2V}\left(\theta+A^0\right)^2\,,
\end{equation}
with the six-dimensional base space of
\begin{equation}
e^{-V}ds^2(B_6)\,=\,\frac{1}{8}e^{2U_1}ds^2_{\mathbb{CP}^1}+\frac{1}{8}e^{2U_2}ds^2_{\mathbb{CP}^1}+\frac{1}{8}e^{2U_3}ds^2_{S^2}\,.
\end{equation}
The warp factors are
\begin{equation}
u_1\,=\,U_1+\frac{1}{2}V\,, \qquad u_2\,=\,U_2+\frac{1}{2}V\,, \qquad u_3\,=\,U_3+\frac{1}{2}V\,, \qquad \phi\,=\,-U_1-U_2-U_3\,.
\end{equation}
The K\"ahler potential is
\begin{equation}
\mathcal{K}\,=\,\frac{1}{6}\mathcal{K}_{ijk}v^iv^jv^k\,,
\end{equation}
and for $Q^{1,1,1}$ solutions,
\begin{equation}
\mathcal{K}_{123}\,=\,1\,.
\end{equation}
There is a relation which we employ later,
\begin{equation} \label{phivk}
e^{2\phi}\,=\,e^{3V}\mathcal{K}^{-1}\,.
\end{equation}

The four-form flux is given by
\begin{align}
G_4\,=\,dA_3+G_4^{\text{flux}}=&\,H_4+dB\wedge\left(\theta+A^0\right)+H_2^i\wedge\omega_i+Db^i\wedge\omega_i\wedge\left(\theta+A^0\right) \notag \\
+&D\xi^A\wedge\alpha_A-D\tilde{\xi}_A\wedge\beta^A+\chi_i\tilde{\omega}^i \notag \\
+&\left[\left(b^IQ^I+\mathbb{U}\xi\right)^A\alpha_A-\left(b^IQ_I+\mathbb{U}\xi\right)_A\beta^A\right]\wedge\left(\theta+A^0\right)\,,
\end{align}
where
\begin{equation}
H_4\,=\,\mathcal{K}^{-1}e^{4\phi}\left(b^I\mathcal{E}_I+\frac{1}{2}\mathcal{K}_{ijk}m^ib^jb^k\right)*1\,,
\end{equation}
and
\begin{equation}
\mathcal{E}_I\,=\,e_I+Q_I^T\mathbb{C}\xi-\frac{1}{2}\delta^0_I\xi^T\mathbb{CU}\xi\,.
\end{equation}
On $SE_7$ there exists a set of real differential forms, an one-form, $\theta$, $n_V$ two-forms, $\omega_i$, 2$n_H$ three-forms, $\alpha_A$, $\beta^A$, $n_V$ four-forms, $\tilde{\omega}^i$, and a six form, $\tilde{\omega}^0$ where $n_V$ and $n_H$ are the numbers of vector and hypermultiplets. Four-form fluxes, ($p^A$, $q_A$), and geometric fluxes, ($m_i\,^A$, $e_{iA}$), and ($v^A\,_B$, $t^{AB}$, $s_{AB}$, $u_A\,^B$), define matrices of
\begin{equation}
Q_I\,=\,\left(
\begin{array}{ll}
 p^A & m_i\,^A \\
 q_A & e_{iA}
\end{array}
\right)\,, \qquad
\mathbb{U}\,=\,\left(
\begin{array}{ll}
 v^A\,_B & t^{AB} \\
 s_{AB} & u_A\,^B
\end{array}
\right)\,,
\end{equation}
where we have
\begin{equation}
v^A\,_B\,=\,-u_B\,^A\,.
\end{equation}
The $Sp(2n_H,\mathbb{R})$ matrix is defined to be
\begin{equation}
\mathbb{C}\,=\,\left(
\begin{array}{ll}
 \,\,\,\,\,\, 0 & \delta_A\,^B \\
 -\delta^A\,_B & \,\,\,\, 0
\end{array}
\right)\,,
\end{equation}
and the scalar fields from hypermultiplets parametrize
\begin{equation}
\xi\,=\,\left(
\begin{array}{ll}
 \xi^A \\
 \tilde{\xi}_A
\end{array}
\right)\,.
\end{equation}
Some parameters are introduced to be
\begin{equation}
e_I\,=\,\left(e_0, e_i\right)\,, \qquad m^I\,=\,\left(0, m^i\right)\,, \qquad b^I\,=\,\left(1, b^i\right)\,,
\end{equation}
with a choice of
\begin{equation}
e_i\,=\,0\,.
\end{equation}
The constant, $e_0$, is a dualization of three-form potential on four-dimensional external spacetime. 

When the Betti number, $b_3\,=\,0$, which is the case we consider, we have the four-form and geometric fluxes to be
\begin{equation}
e_i\,=\,0\,, \qquad p^A\,=\,0\,, \qquad q_A\,=\,0\,.
\end{equation}
For $Q^{1,1,1}$ and $M^{1,1,1}$ solutions, we have the geometric fluxes of
\begin{equation}
e_{iA}\,=\,0\,, \qquad m_i\,^A\,=\,0\,,
\end{equation}
and
\begin{equation}
u_A\,^B\,=\,0\,, \qquad s_{00}\,=\,-4\,, \qquad t^{00}\,=\,4\,.
\end{equation}
Also for $Q^{1,1,1}$ and $M^{1,1,1}$ solutions, we have
\begin{equation}
m^1\,=\,m^2\,=\,m^3\,=\,2\,.
\end{equation}

\subsection{M2-brane probe}

At the $AdS_4$ fixed points, we have
\begin{equation}
ds^2_4\,=\,e^{2A}\left(-dx_0^2+dx_1^2+dx_2^2\right)+dr^2\,,
\end{equation}
where
\begin{equation}
A\,=\,\frac{r}{l}\,,
\end{equation}
and $l$ is the radius of $AdS_4$. 
The relevant part of four-form flux is
\begin{equation}
G_4\,=\,\mathcal{K}^{-1}e^{4\phi}\left(e_0+2\left((\xi^0)^2+(\tilde{\xi}_0)^2\right)+2b^1b^2+2b^2b^3+2b^3b^1\right)vol_4\,.
\end{equation}
Thus we obtain that the three-form potential is
\begin{equation}
A_3\,=\,\frac{l}{3}e^{3A}\mathcal{K}^{-1}e^{4\phi}\left(e_0+2\left((\xi^0)^2+(\tilde{\xi}_0)^2\right)+2b^1b^2+2b^2b^3+2b^3b^1\right)dx_0\wedge{d}x_1\wedge{d}x_2\,.
\end{equation}

We partition the spacetime coordinates,
\begin{equation}
x^a\,=\,\{x_0,x_1,x_2\}\,, \qquad y^m\,=\,\{r,\ldots\}\,,
\end{equation}
and choose the static gauge,
\begin{equation}
x_0\,=\,t\,=\,\eta^0\,,\qquad x^a\,=\,\eta^a\,, \qquad y^m\,=\,y^m(t)\,,
\end{equation}
where $\eta^a$ are the worldvolume coordinates. The pull-back of the metric is
\begin{equation}
\tilde{G}_{ab}\,=\,G_{\mu\nu}\frac{\partial{x}^\mu}{\partial\eta^a}\frac{\partial{x}^\nu}{\partial\eta^b}\,.
\end{equation}

Now we study the worldvolume action of the M2-branes which is given by a sum of DBI and WZ terms. If the probe branes move slowly, the worldvolume action is
\begin{align}
S\,=&\,-\int{d}^3\eta\sqrt{-\text{det}(\tilde{G})}+\int\tilde{A}_3 \notag \\
=&\,-\int{d}^3\eta\left(e^{3V+3A}\mathcal{K}^{-3/2}-\frac{1}{2}e^{V+A}\mathcal{K}^{-1/2}G_{mn}\dot{y}^m\dot{y}^n+\ldots\right) \notag \\
&\,+\int\frac{l}{3}{e}^{3A}\mathcal{K}^{-1}e^{4\phi}\left(e_0+2\left((\xi^0)^2+(\tilde{\xi}_0)^2\right)+2b^1b^2+2b^2b^3+2b^3b^1\right)\,{d}x_0\wedge{d}x_1\wedge{d}x_2\,,
\end{align}
where $\tilde{A}_3$ is the pull-back of the three-form potential. Then the worldvolume action reduces to
\begin{equation}
S\,=\,\int{d}^3\eta\left(K-\mathcal{V}\right)\,,
\end{equation}
where the kinetic and the potential terms are
\begin{align}
K\,=&\,\frac{1}{2}e^{V+A}\mathcal{K}^{-1/2}G_{mn}\dot{y}^m\dot{y}^n+\ldots\,, \notag \\
\mathcal{V}\,=&\,e^{3A}\left(e^{3V}\mathcal{K}^{-3/2}-\frac{l}{3}e^{6V}\mathcal{K}^{-3}\left(e_0+2\left((\xi^0)^2+(\tilde{\xi}_0)^2\right)+2b^1b^2+2b^2b^3+2b^3b^1\right)\right)\,,
\end{align}
where we employed the relation in \eqref{phivk}. For the supersymmetric fixed point in \eqref{susyfp}, we obtain
\begin{equation}
e^{-3A}\mathcal{V}|_{\text{SUSY}}\,=\,0\,.
\end{equation}
For the non-supersymmetric fixed point in \eqref{nonsusyfp}, we obtain
\begin{equation}
e^{-3A}\mathcal{V}|_{\text{non-SUSY}}\,=\,\frac{1}{\sqrt{15}\,a^{21/4}}\left(\frac{7}{2}\right)^{3/4}+\frac{7\sqrt{14}\,l^{9/4}}{45\,a^{15/2}}\,,
\end{equation}
where the free parameter, $a>0$, and $l$ is the radius of $AdS_4$. Thus the force acting on the probe M2-brane, $d\mathcal{V}/dr$, is positive and attractive. We conclude that both supersymmetric and non-supersymmetric vacua are brane-jet stable. They are also known to be BF stable within the truncation of \cite{Cassani:2012pj}.

So far we have considered the solutions of $AdS_4\,\times\,Q^{1,1,1}$. The analysis for the $AdS_4\,\times\,M^{1,1,1}$ solutions goes parallel by identifying $t^3\,=\,t^1$ and $U_2\,=\,U_1$. Therefore, the vacua from $M^{1,1,1}$ are also BF and brane-jet stable.

\section{Conclusions}

In this paper, we have examined the brane-jet instabilities of diverse non-supersymmetric $AdS$ solutions: the Janus solutions of type IIB supergravity and the $AdS_4$ vacua from eleven-dimensional supergravity on Sasaki-Einstein manifolds. We showed that all $AdS$ vacua we considered are brane-jet stable. Thus, among those solutions, the Janus, the skew-whipped Freund-Rubin, and the $AdS_4$ vacua from $Q^{1,1,1} $ and $M^{1,1,1}$ manifolds are perturbatively stable within subsectors of truncations and also brane-jet stable. They make candidates of the counter-example to the swampland conjecture on the instability of non-supersymmetric $AdS$ vacua.

In the analysis we have extended the application of brane-jets to the $AdS$ vacua from curved domain walls. Unlike the flat domain walls where the worldvolume of probe brane is on Minkowski spacetimes, the worldvolume is on the curved spacetimes for curved domain walls. Depending on the geometry of domain walls, identical solutions display different behaviours of brane-jets. For the Janus solutions we considered, the worldvolume was on $AdS_4$ instead of Mink$_4$. This suggests that we could apply the brane-jet analysis to various $AdS$ vacua from more general geometries.

Unlike the usual solutions studied for brane-jets, the solutions from Sasaki-Einstein manifolds are not warped but direct products of $AdS_4$ and internal manifolds. Thus the probe brane potentials were independent of the internal coordinates and were constant. There are similar examples of direct product solutions studied previously for brane-jets: the $SO(7)^-$ solution of eleven-dimensional supergravity was proven to be brane-jet stable in \cite{Bena:2020xxb} and the $G_2$ and $SO(7)$ solutions of massive type IIA supergravity were brane-jet stable and unstable, respectively, in \cite{Guarino:2020jwv}.

If the $AdS$ vacua we have found to be both perturbatively and brane-jet stable are truly stable, it would be most interesting to establish the precise AdS/CFT correspondence, \cite{Maldacena:1997re}, in the non-supersymmetric setting. For the skew-whiffed Freund-Rubin solutions, possible dual field theories were studied in \cite{Berkooz:1998qp, Forcella:2011pp}. However, there is always a possibility of new instabilities we have not yet discovered. Some potential instabilities from global singlet marginal operators and tunneling into bubble of nothing were discussed in \cite{Berkooz:1998qp, Murugan:2016aty}.

\bigskip
\leftline{\bf Acknowledgements}
\noindent We would like to thank Krzysztof Pilch and Oscar Varela for reading a draft of the manuscript. We are happy to thank an anonymous referee for an instructive comment on \eqref{321}. This research was supported by the National Research Foundation of Korea under the grant  NRF-2019R1I1A1A01060811.

\appendix
\section{The $AdS_5\times{S}^5$ solution of type IIB supergravity}
\renewcommand{\theequation}{A.\arabic{equation}}
\setcounter{equation}{0} 

We consider the normalization of the $AdS_5\times{S}^5$ solution of type IIB supergravity. See $e.g.$, \cite{Duff:1998us}. The metric and the self-dual five-form flux are given, respectively, by
\begin{align}
ds^2\,=&\,ds_{AdS_5}^2+\frac{1}{m^2}ds_{S^5}^2\,, \notag \\
F_{(5)}\,=&\,4m\,\text{vol}_{AdS_5}+4m\,\text{vol}_{S^5}\,,
\end{align}
where $m$ is a free parameter and we employed the metric and the volume form with unit radius. Then the Riemann tensors of $AdS_5$ and $S^5$ are given by
\begin{equation}
R_{\mu\nu}\,=\,-4m^2g_{\mu\nu}\,, \qquad R_{mn}\,=\,4m^2g_{mn}\,.
\end{equation}
In the main body of the manuscript, we have set $m=1$.

\vspace{1.5cm}

\bibliographystyle{JHEP}
\bibliography{20210222}

\end{document}